\documentclass[doublecol]{epl2} 

\usepackage{amsmath}
\usepackage{graphicx,graphics}

\newcommand\rhopa{{\varrho}}
\newcommand\rhotot{{\boldsymbol{\rho}}}

\title{Electronic phase coherence versus dissipation 
in solid-state quantum devices: Two approximations are better than one}
\shorttitle{} 

\author{R. C. Iotti \and F. Rossi\thanks{E-mail: \email{fausto.rossi@polito.it}}}
\shortauthor{}
\institute{                    
Department of Applied Science and Technology, Politecnico di Torino, C.so Duca degli Abruzzi 24, 10129 Torino, Italy
}

\abstract{In the microscopic modeling of new-generation electronic quantum nanodevices a variety of simulation strategies have been proposed and employed. Aim of this Letter is to point out virtues versus intrinsic limitations of non-Markovian density-matrix approaches; we shall show that the usual mean-field treatment may lead to highly unphysical results, like negative distribution probabilities and non-dissipative behaviours, which are particularly severe in zero-dimensional electronic systems coupled to dispersionless phonon modes.
This is in striking contrast with Markovian treatments, where a proper combination of adiabatic limit and mean-field schemes guarantees a physically acceptable solution; as a result, the unusual conclusion is that two approximations are better than one.
}

\begin{document}

\maketitle

Solid-state electronic quantum devices currently involve a wide variety of materials and encoding strategies, namely quantum-information schemes based on charge and/or spin degrees of freedom, in both semiconducting and superconducting systems (see, e.g., \cite{Fazio05a,b-Rossi05,b-Benson09}).
In particular, artificially tailored as well as self-assembled nanostructures form the leading edge of semiconductor science and technology; the design of state-of-the-art optoelectronic devices heavily exploits the principles of band-gap engineering (see, e.g., \cite{b-Capasso11}), achieved by confining charge carriers in spatial regions comparable to their de Broglie wavelengths. 
This, together with the progressive reduction of the typical time-scales involved, pushes device miniaturization toward limits where the application of the traditional Boltzmann transport theory becomes questionable, and more rigorous quantum-transport approaches are imperative. The latter can be qualitatively subdivided into two main classes: double-time approaches based on the nonequilibrium Green's function technique (see, e.g., \cite{b-Haug07,b-Datta05}) and single-time approaches based on the density-matrix theory (see, e.g., \cite{b-Haug04,b-Rossi11}).

Regardless of the specific method, the derivation of effective kinetic equations for the subsystem of interest ---namely the electronic gas in a quantum device--- may involve one or more of the following three key steps:\footnote{A relevant exception is the so-called ``dynamics controlled truncation'' introduced by Axt and Stahl (see, e.g., \cite{Axt98a}) based on an expansion in powers of the exciting laser field.} (i) mean-field approximation, (ii) adiabatic or Markov limit, and (iii) semiclassical or diagonal limit.
As a result of all these three approximations one gets the usual Boltzmann transport equation; the latter, when applicable, constitutes a robust and reliable particle-like description in purely stochastic terms, providing physically acceptable results.
In contrast, the combination of only the first two approximation schemes, namely mean-field treatment and adiabatic limit, allows one to derive so-called Markovian scattering superoperators, whose action may lead to unphysical results; indeed, the choice of the adiabatic decoupling strategy is definitely not unique, and, in general, the positive-definite character of the density-matrix operator may be violated. To overcome this serious limitation, a few years ago an alternative Markov procedure has been proposed \cite{Taj09b}, allowing for a microscopic derivation of Lindblad-like scattering superoperators and thus preserving the positive-definite nature of the many-electron subsystem. More recently, such alternative Markov scheme combined with the conventional mean-field approximation has allowed for the derivation of a positive-definite nonlinear equation for the single-particle density matrix \cite{Rosati14e}.

When the system-environment coupling  becomes  strong and/or the excitation timescale is extremely short, Markovian approaches are known to be unreliable, and memory effects have to be taken into account via so-called quantum-kinetic schemes (see, e.g., \cite{b-Haug07,b-Bonitz98}); indeed, stimulated by the pioneering works by Haug and coworkers \cite{TranThoai93a}, as well as Kuhn and coworkers \cite{Schilp94a}, over the last decade several groups have routinely employed such non-Markovian techniques to investigate a large variety of ultrafast phenomena in semiconductors (see, e.g., \cite{Butscher05a,Vu06a,Gartner06a,Rozbicki08a,GrodeckaGrad10a,Papenkort12a,Cygorek14a}).
In this case the only approximation involved is the mean-field factorization scheme, and, in general, it is hard to draw conclusions about the physical versus non-physical nature of the resulting carrier dynamics.

Aim of this Letter is to point out advantages versus intrinsic limitations of non-Markovian density-matrix approaches. More specifically, we shall show that the usual mean-field treatment employed to derive quantum-kinetic equations may lead to highly unphysical results, like negative distribution functions and non-dissipative carrier-phonon couplings; by means of a simple two-level model, we shall show that such limitations are particularly severe in zero-dimensional (0D) electronic systems ---like quantum-dot-based nanodevices--- coupled to dispersionless phonon modes.
Such a behaviour is in striking contrast with the case of Markovian treatments, where a proper combination of adiabatic limit and mean-field approximation (see, e.g., \cite{Rosati14e}) guarantees a physically acceptable solution; as a result, in this case, the unusual conclusion is that two approximations are better than one.

In order to describe phonon-induced carrier dissipation and decoherence in state-of-the-art quantum nanodevices, let us consider the following many-body Hamiltonian:
\begin{eqnarray}\label{H}
\hat H 
&=& 
\sum_\alpha \epsilon_\alpha \hat c^\dagger_\alpha \hat c^{ }_\alpha
+
\sum_{\mathbf{q}} \epsilon_{\mathbf{q}} \hat b^\dagger_{\mathbf{q}} \hat b^{ }_{\mathbf{q}} \nonumber \\
&+&
\sum_{\alpha\alpha',\mathbf{q}}
\left( 
g^{\mathbf{q},-}_{\alpha\alpha'}
\hat c^\dagger_{\alpha} 
\hat b^{ }_{\mathbf{q}} 
\hat c^{ }_{\alpha'} 
+
g^{\mathbf{q},+}_{\alpha\alpha'}
\hat c^\dagger_{\alpha'} 
\hat b^\dagger_{\mathbf{q}} 
\hat c^{ }_{\alpha}
\right)\ .
\end{eqnarray}
Here, the first two terms describe the noninteracting carrier and phonon systems in terms of electronic states $\alpha$ and phonons of wavevector $\mathbf{q}$ (with energies $\epsilon_\alpha$ and $\epsilon_{\mathbf{q}}$, respectively), while the last one accounts for carrier-phonon interaction. More specifically, $-$ ($+$) refers to phonon absorption (emission), while the explicit form of the coupling matrix elements 
$
g^{\mathbf{q},\pm}_{\alpha\alpha'}
$ 
depends on the particular
phonon branch (acoustic, optical, etc.) as well as on the coupling
mechanism considered (deformation potential, polar coupling, etc.).

Following the general quantum-kinetic approach originally proposed in \cite{Schilp94a} and reviewed in \cite{Rossi02b}, let us introduce the single-particle density matrix
$\rho_{\alpha_1\alpha_2} = \langle \hat c^\dagger_{\alpha_2} \hat c^{ }_{\alpha_1} \rangle$,
the so-called phonon-assisted density matrices
$\rhopa^{\mathbf{q}}_{\alpha_1\alpha_2} = \langle \hat c^\dagger_{\alpha_2} \hat b^{ }_{\mathbf{q}} \hat c^{ }_{\alpha_1} \rangle$,
and the phonon distribution
$n_{\mathbf{q}} = \langle \hat b^\dagger_{\mathbf{q}} \hat b^{ }_{\mathbf{q}} \rangle$.
By employing a well-established correlation-expansion procedure, combined with a proper mean-field treatment of the carrier-phonon system, and neglecting various renormalization terms as well as coherent-phonon contributions, one gets the following set of coupled kinetic equations:\footnote{Here, the kinetic equation for the phonon distribution $n^{ }_{\mathbf{q}}$ describes how the electron gas may drive the phonon system out of the thermal-equilibrium distribution $n^\circ_{\mathbf{q}}$. 
Indeed, the impact of the so-called hot phonons derives from the interplay between phonon heating due to phonon emission by nonequilibrium electrons and phonon thermalization via phonon-phonon interaction; here the latter is accounted for within a relaxation-time model in terms of a constant rate $\Gamma_{\rm p}$.
Such hot-phonon effects are known to play an important role in the design and optimization of quantum-cascade devices (see, e.g., \cite{Iotti13a}).}
\begin{eqnarray}\label{QKE1}
\frac{d\rho_{\alpha_1\alpha_2}}{dt}
&=&
\frac{1}{\imath\hbar} \left(\epsilon_{\alpha_1} - \epsilon_{\alpha_2}\right) \rho_{\alpha_1\alpha_2}
\nonumber \\
&+&
\frac{1}{\imath\hbar} \sum_{\alpha_3,\mathbf{q}} 
\left(
g^{\mathbf{q},-}_{\alpha_1\alpha_3} \rhopa^{\mathbf{q}}_{\alpha_3\alpha_2}
+ 
g^{\mathbf{q},+}_{\alpha_3\alpha_1} \rhopa^{\mathbf{q} *}_{\alpha_2\alpha_3}
\right)\,+\,\textrm{H.c.}
\nonumber \\
\frac{d\rhopa^{\mathbf{q}}_{\alpha_1\alpha_2}}{dt}
&=&
\frac{1}{\imath\hbar} 
\left(\epsilon^*_{\alpha_1} - \epsilon^{ }_{\alpha_2} + \epsilon_\mathbf{q}\right) 
\rhopa^\mathbf{q}_{\alpha_1\alpha_2}
\nonumber \\
&+&
\frac{1}{\imath\hbar} \sum_{\alpha_3\alpha_4} 
g^{\mathbf{q},+}_{\alpha_4\alpha_3} 
\left(n_{\mathbf{q}} + 1\right)
\rho_{\alpha_4\alpha_2} 
\left(\delta_{\alpha_1\alpha_3} - \rho_{\alpha_1\alpha_3}\right)
\nonumber \\
&-& 
\frac{1}{\imath\hbar} \sum_{\alpha_3\alpha_4} 
g^{\mathbf{q},+}_{\alpha_4\alpha_3} 
n_{\mathbf{q}} 
\rho_{\alpha_1\alpha_3}
\left(\delta_{\alpha_4\alpha_2} - \rho_{\alpha_4\alpha_2}\right)
\nonumber \\
\frac{d n_{\mathbf{q}}}{dt}
=& 
-&\Gamma_{\rm p}\left(n^{ }_{\mathbf{q}} - n^\circ_{\mathbf{q}}\right)
- \frac{1}{\imath\hbar} \sum_{\alpha_1\alpha_2} 
g^{\mathbf{q},-}_{\alpha_1\alpha_2} \rhopa^{\mathbf{q}}_{\alpha_2\alpha_1}
\,+\,\textrm{c.c.}\ .
\end{eqnarray}
In the low-density limit, i.e., $(\delta_{\alpha\alpha'} - \rho_{\alpha\alpha'}) \to \delta_{\alpha\alpha'}$, hot-phonon effects vanish, i.e., $n^{ }_{\mathbf{q}} \to n^\circ_{\mathbf{q}}$, and the original set of quantum-kinetic equations in (\ref{QKE1}) reduces to:\begin{eqnarray}\label{QKE2}
\frac{\upd\rho_{\alpha_1\alpha_2}}{\upd t}
&=&
\frac{1}{\imath\hbar} \left(\epsilon_{\alpha_1} - \epsilon_{\alpha_2}\right) \rho_{\alpha_1\alpha_2}
\nonumber \\
&+&
\frac{1}{\imath\hbar} \sum_{\alpha_3,\mathbf{q}} 
\left(
g^{\mathbf{q},-}_{\alpha_1\alpha_3} \rhopa^{\mathbf{q}}_{\alpha_3\alpha_2}
+ 
g^{\mathbf{q},+}_{\alpha_3\alpha_1} \rhopa^{\mathbf{q} *}_{\alpha_2\alpha_3}
\right)\,+\,\textrm{H.c.}
\nonumber \\
\frac{\upd\rhopa^{\mathbf{q}}_{\alpha_1\alpha_2}}{\upd t}
&=&
\frac{1}{\imath\hbar} 
\left(\epsilon^*_{\alpha_1} - \epsilon^{ }_{\alpha_2} + \epsilon_\mathbf{q}\right) 
\rhopa^\mathbf{q}_{\alpha_1\alpha_2}
\nonumber \\
&+&
\frac{1}{\imath\hbar} \sum_{\alpha_3} \!
\left[
g^{\mathbf{q},+}_{\alpha_3\alpha_1} \!
\left(n_{\mathbf{q}} \!+\! 1\right) 
\rho_{\alpha_3\alpha_2} 
\!-
g^{\mathbf{q},+}_{\alpha_2\alpha_3} 
n_{\mathbf{q}} 
\rho_{\alpha_1\alpha_3} 
\right] \, .
\end{eqnarray}
Applying to the above coupled equations the alternative Markov limit proposed in \cite{Taj09b}, the phonon-assisted density matrix $\rhopa^{\mathbf{q}}_{\alpha_1\alpha_2}$ can be adiabatically eliminated, leading to the following Lindblad-type Markovian dynamics:
\begin{eqnarray}\label{LBE}
\frac{\upd\rho_{\alpha_1\alpha_2}}{\upd t}
&=&
\frac{1}{\imath\hbar} \left(\epsilon_{\alpha_1} - \epsilon_{\alpha_2}\right) \rho_{\alpha_1\alpha_2}
\nonumber \\
&+&
\frac{1}{2} \sum_{\alpha'_1\alpha'_2}
\mathcal{P}^{ }_{\alpha_1\alpha_2,\alpha'_1\alpha'_2} \rho_{\alpha'_1\alpha'_2}
\,+\,{\rm H.c.}\nonumber \\
&-&
\frac{1}{2} \sum_{\alpha'_1\alpha'_2}
\mathcal{P}^{*}_{\alpha'_1\alpha'_1,\alpha_1\alpha'_2}
\rho_{\alpha'_2\alpha_2} 
\,+\,{\rm H.c.}
\end{eqnarray}
with generalized scattering rates
\begin{equation}\label{calP}
\mathcal{P}^{ }_{\alpha_1\alpha_2,\alpha'_1\alpha'_2} = \sum_{\mathbf{q},\pm}
A^{\mathbf{q},\pm}_{\alpha_1\alpha'_1} 
A^{\mathbf{q},\pm *}_{\alpha_2\alpha'_2}\ ,
\end{equation}
where
\begin{equation}\label{A}
A^{\mathbf{q},\pm}_{\alpha\alpha'} =
\sqrt{
\frac{
2 \pi \left(n^\circ_{\mathbf{q}}+\frac{1}{2}\pm\frac{1}{2}\right)
} 
{
\hbar
}
}
g^{\mathbf{q},\pm}_{\alpha\alpha'} D^{\mathbf{q},\pm}_{\alpha\alpha'}
\end{equation}
and
\begin{equation}
D^{\mathbf{q},\pm}_{\alpha\alpha'} =
\lim_{\overline{\epsilon} \to 0}
\frac{
e^{-\left(
\frac{\epsilon_\alpha-\epsilon_{\alpha'} \pm \epsilon_{\mathbf{q}} 
}
{ 
2 \overline{\epsilon}}\right)^2 }
}
{ 
\left(2\pi \overline{\epsilon}^2\right)^{\frac{1}{4}}
} \ .
\end{equation}
The diagonal elements 
$P_{\alpha \to \alpha'} \equiv \mathcal{P}^{ }_{\alpha'\alpha',\alpha\alpha}$ 
of the generalized rates (\ref{calP}) coincide with the usual Fermi's-golden-rule prescription of the semiclassical theory, and correspond to an electron-phonon inverse lifetime
$\Gamma_\alpha = \sum_{\alpha'} P_{\alpha \to \alpha'}$.

In order to point out virtues versus intrinsic limitations of the quantum-kinetic treatment in Eq.~(\ref{QKE2}) [compared to its Markovian counterpart in Eq.~(\ref{LBE})], 
let us consider an electronic two-level system ($\alpha \equiv \{a,b\}$) characterized by an energy splitting $\Delta_{\rm c} = \epsilon_b -\epsilon_a$ and described by a two-by-two density matrix in terms of the ground- and excited-level populations $f_a = \rho_{aa}$ and $f_b = \rho_{bb}$ as well as of the interlevel polarization $p = \rho^{ }_{ba} = \rho^*_{ab}$.
Moreover, regardless of the specific phonon mode considered, we shall adopt $\mathbf{q}$-independent coupling matrices of the form:
$g^{\mathbf{q},\pm}_{\alpha\alpha'} = g(1-\delta_{\alpha\alpha'})$.
In all the following simulated experiments, we have chosen as initial condition of the electronic system a maximally coherent state 
$\rho_{\alpha_1\alpha_2}(0) = \frac{1}{2}$, 
setting to zero all phonon-assisted density matrices, i.e., 
$\rhopa^{\mathbf{q}}_{\alpha_1\alpha_2}(0) = 0$.\footnote{This choice is however not binding, since analogous features would emerge also with more general (and even diagonal) initial electronic states.}

\begin{figure}
\centerline{\includegraphics[width=.4\textwidth]{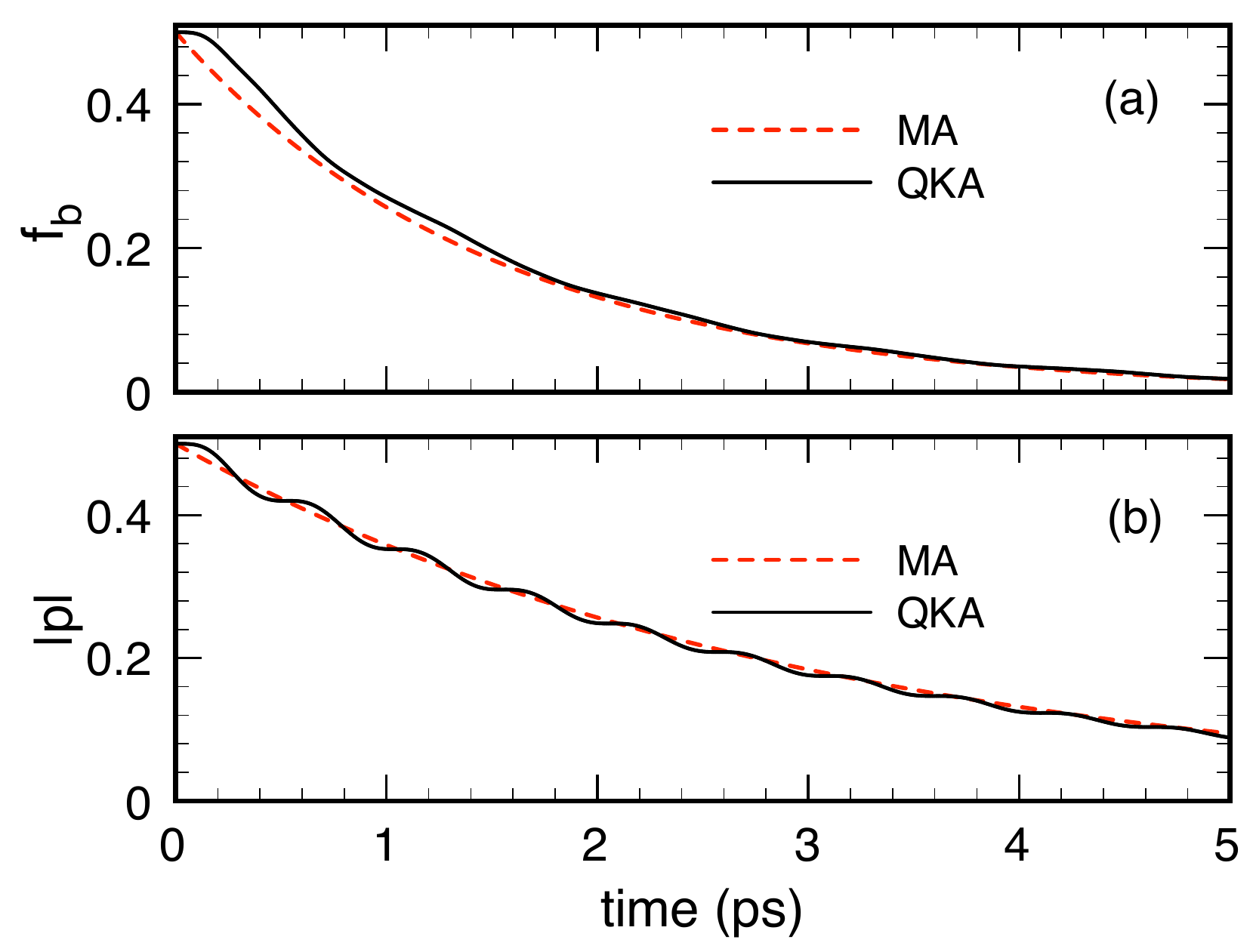}}
\caption{
Energy dissipation and decoherence for an electronic two-level system, induced by an acoustic-like phonon mode in the low-temperature limit and for weak-coupling conditions.
(a): Excited-level population $f_b$ and (b): interlevel-polarization modulus $|p|$ as a function of time, obtained via the quantum-kinetic approach in Eq.~(\ref{QKE2}) (solid curves) and the Markovian approach in Eq.~(\ref{LBE}) (dashed curves).
}
\label{Fig1}
\end{figure}

\begin{figure}
\centerline{\includegraphics[width=.4\textwidth]{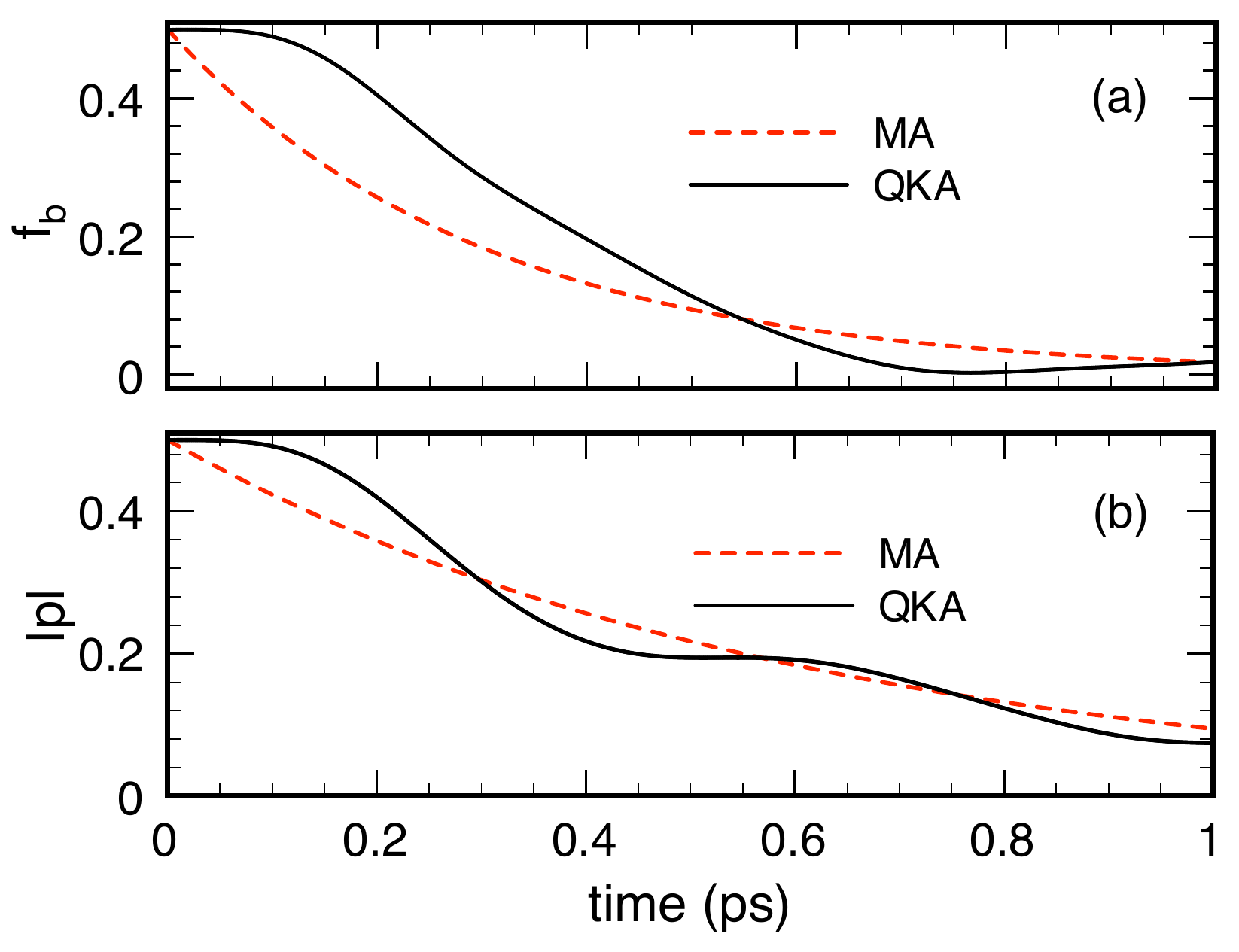}}
\caption{
Same as in Fig.~\ref{Fig1} but in the strong-coupling regime.
}
\label{Fig2}
\end{figure}

As a starting point, let us investigate energy dissipation and decoherence induced on our two-level system by an acoustic-like phonon mode, whose linear-dispersion bandwidth $0 \le \epsilon_{\mathbf{q}} \le \Delta_{\rm p}$ is much greater than the interlevel splitting $\Delta_{\rm c}$.
Figure~\ref{Fig1} shows a comparison between the results obtained within the Markovian approach (MA) and the quantum-kinetic approach (QKA) in the low-temperature limit 
($n^\circ_{\mathbf{q}} = 0$), for $\Delta_{\rm c} = 4$\,meV and for weak-coupling conditions; to this end, the coupling coefficient $g$ has been chosen such to produce a semiclassical scattering time $P^{-1}_{b \to a} = 1.5$\,ps, corresponding to an effective interlevel coupling energy $\Delta_{\rm cp}$ of about $0.4$\,meV.
As expected, in this weak-coupling regime the quantum-kinetic results exhibit very small deviations with respect to the Markovian ones; in particular, the initial time derivative of the excited-level population is equal to zero, a well-known fingerprint of non-Markovian treatments (see also Fig.~\ref{Fig2}).

To point out genuine non-Markovian features, we repeat the simulated experiments presented so far increasing the effective interlevel coupling energy $\Delta_{\rm cp}$ by a factor of five. Figure \ref{Fig2} shows the new (strong-coupling) scenario.
As expected, while the Markov results are simply rescaled by a factor of five, the quantum-kinetic dynamics shows significant deviations with respect to its Markovian counterpart. In particular, the transient is strongly affected by so-called energy nonconserving transitions (see, e.g., \cite{b-Rossi11}); this, in addition to the zero-derivative behaviour, gives rise to an initial reduction of energy dissipation and decoherence, which tends to vanish at longer times.

Both the weak-coupling results of Fig.~\ref{Fig1} and the strong-coupling ones of Fig.~\ref{Fig2} clearly show the power and flexibility of the quantum-kinetic treatment in describing a 0D electronic system coupled to acoustic phonon modes; this is the typical phase-coherence versus dissipation dynamics in quantum-dot-based nanodevices (see, e.g., \cite{b-Rossi05}).\footnote{Actually, a proper description of the strong-coupling regime would need to include multiphonon processes as well (see, e.g., \cite{Axt99a}). This is not however the focus of the present Letter, which instead is now going to address criticalities that may arise already in a weak-coupling scenario.}

Let us now move from acoustic- to optical-like phonon modes; to this end, it is imperative to recall that in the presence of a purely 0D electronic system coupled to a dispersionless phonon mode (corresponding to a discrete electron-plus-phonon energy spectrum) the Markov limit is not applicable, and more refined treatments, based on the polaronic picture (see, e.g., \cite{Grange07a} and references therein), are required.
In view of the above, we have considered an optical-like phonon mode, i.e., a mode characterized by a finite bandwidth $\Delta_{\rm p}$ much smaller than the interlevel splitting $\Delta_{\rm c}$ and centered around the electronic interlevel excitation, for which the Markov treatment is still well defined; more specifically, in order to mimic carrier-LO phonon coupling in GaAs-based nanomaterials, we have chosen an interlevel energy splitting $\Delta_{\rm c} = 40$\,meV, assuming a phonon bandwidth $\Delta_{\rm p} = 4$\,meV and a semiclassical scattering time $P^{-1}_{b \to a} = 0.3$\,ps, corresponding to an effective interlevel coupling energy $\Delta_{\rm cp}$ of about $2$\,meV.

\begin{figure}
\centerline{\includegraphics[width=.4\textwidth]{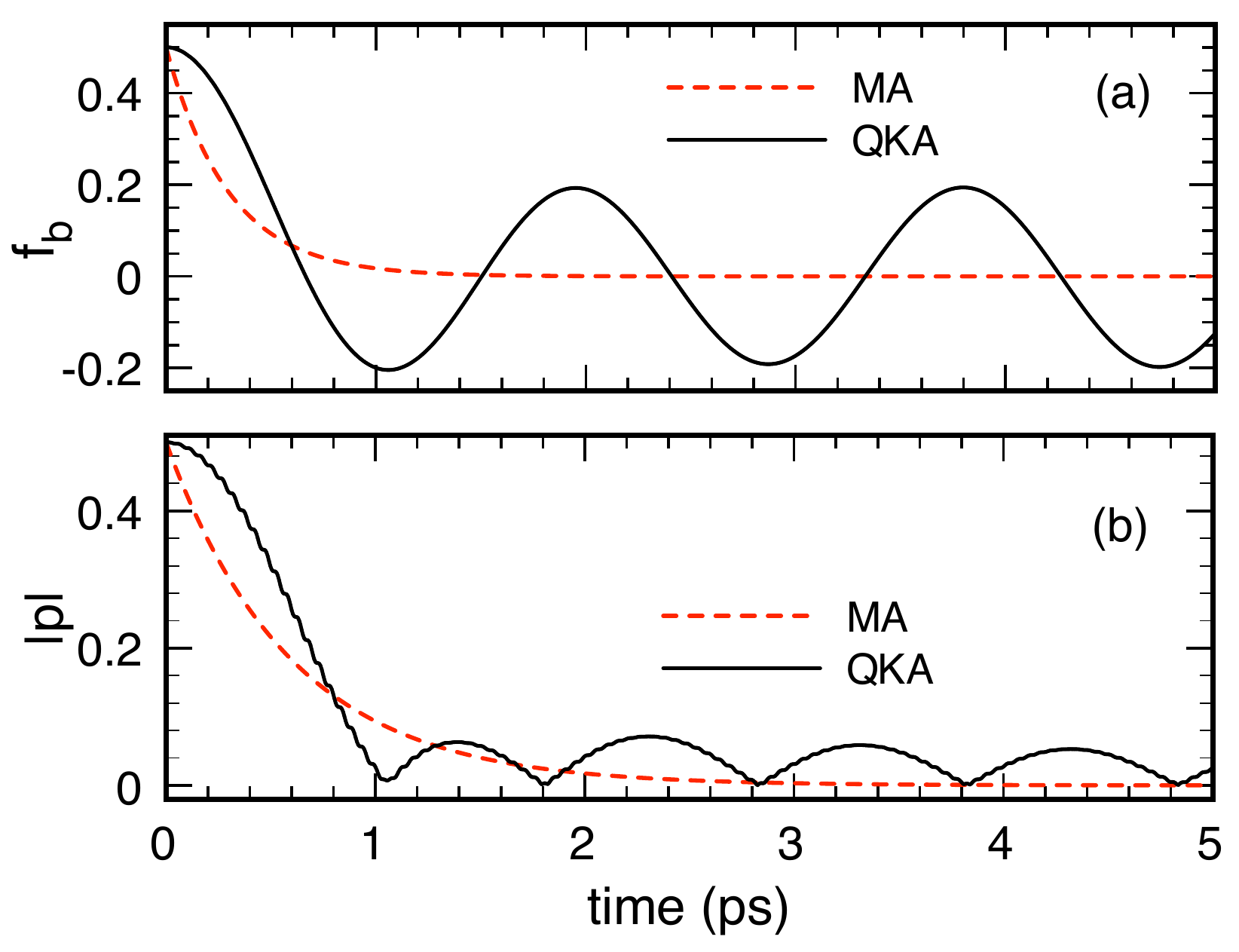}}
\caption{
Same as in Fig.~\ref{Fig1} but for a two-level system with a larger energy splitting,  
weakly coupled to an optical-like phonon mode.
}
\label{Fig3}
\end{figure}

Figure \ref{Fig3} shows again a comparison between Markovian and non-Markovian results in the low-temperature limit for the optical-like phonon mode just described. Here, opposite to the acoustic-phonon results of Fig.~\ref{Fig1}, the quantum-kinetic dynamics is completely different from its Markovian counterpart, in spite of the weak-coupling regime ($\Delta_{\rm c} \simeq 20 \Delta_{\rm cp}$). 
Indeed, while the latter exhibits the usual exponential decay, the non-Markovian results display an almost dissipation-free oscillatory dynamics.
More importantly, the non-Markovian evolution comes out to be totally unphysical, as evidenced by the strongly negative values of the excited-level population $f_b$.

\begin{figure}
\centerline{\includegraphics[width=.4\textwidth]{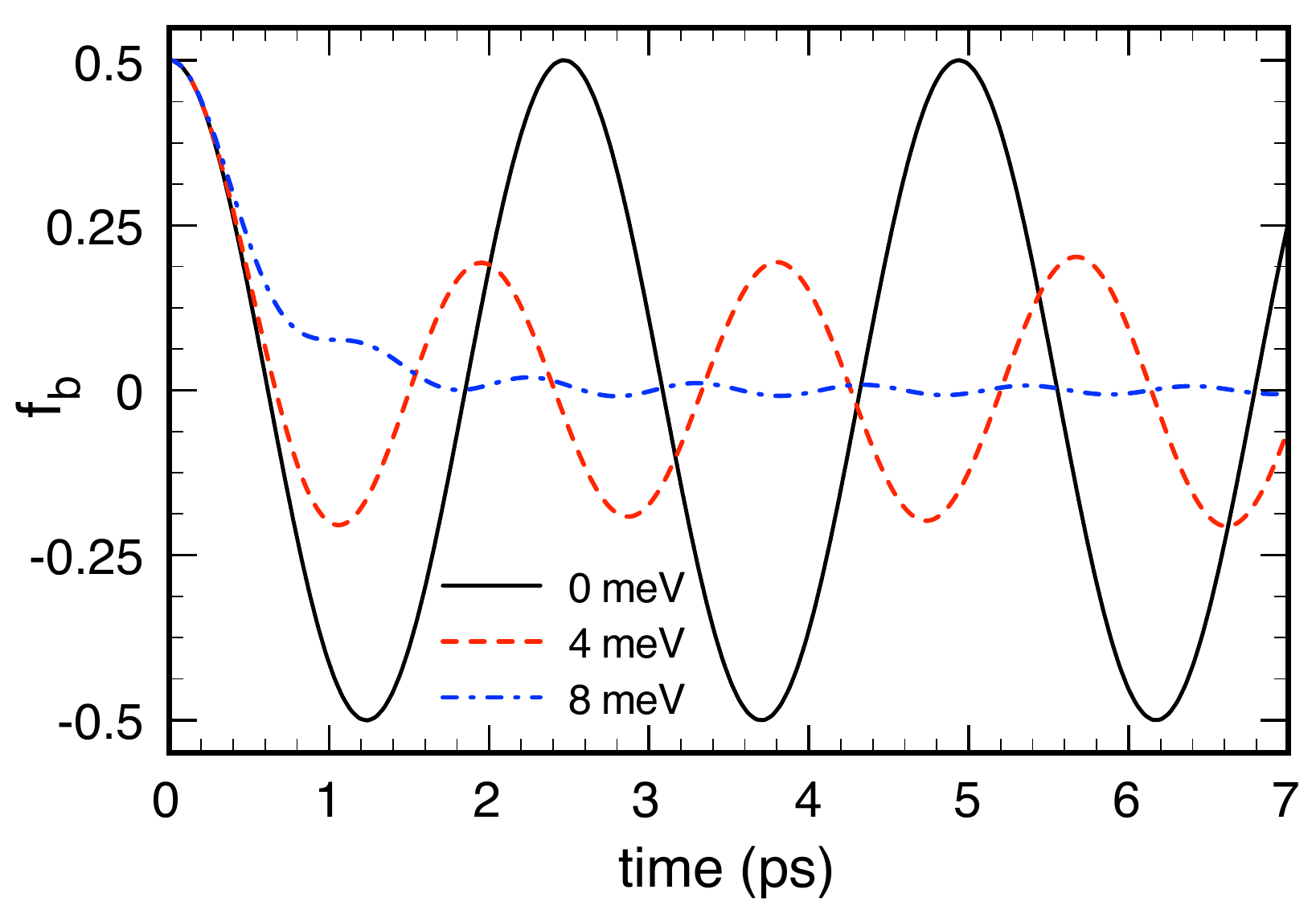}}
\caption{
Quantum-kinetic dynamics of the excited-level population $f_b$ corresponding to the same system considered in Fig.~\ref{Fig3}, for different values of the optical-phonon bandwidth:  $\Delta_{\rm p} = 0$ (solid curve), $4$\,meV (dashed curve) and $8$\,meV (dash-dotted curve).
}
\label{Fig4}
\end{figure}

Comparing the physically sound quantum-kinetic results of Figs.~\ref{Fig1} and \ref{Fig2} with the unphysical ones of Fig.~\ref{Fig3}, one is led to conclude that the critical point is the relatively small value of the optical-phonon bandwidth. This is fully confirmed by the quantum-kinetic analysis presented in Fig.~\ref{Fig4}. Here, the quantum-kinetic dynamics is shown for the three different values $\Delta_{\rm{p}} = 0, 4$ and 8 meV. As one can see, for large values of $\Delta_{\rm p}$ the negative regions are suppressed, showing an increased dissipation effect; in contrast, for vanishing $\Delta_{\rm{p}}$ values one recovers a totally unphysical cosine-like behaviour.
Indeed, in such a dispersionless limit, each phonon $\mathbf{q}$ is characterized by the very same energy, i.e., $\epsilon_{\mathbf{q}} \to \epsilon_b -\epsilon_a$, and the effect of the whole phonon system can be described via a single phonon $\overline{\mathbf{q}}$ resonantly coupled to the two-level system via an effective constant $\overline{g}$. 
More specifically, adopting the usual rotating-wave approximation, the original quantum-kinetic equations (\ref{QKE2}) applied to our two-level system in this limit can be solved analytically: one gets
\begin{equation}\label{as}
f_b(t) 
= f_b(0) \cos\left(\sqrt{2} \, \overline{\omega} t\right)\ , 
\end{equation}
where $\overline{\omega} = \overline{g}/\hbar$, in complete agreement with the $\Delta_{\rm p} \rightarrow 0$ behaviour of Fig.~\ref{Fig4}.

\begin{figure}
\centerline{\includegraphics[width=.4\textwidth]{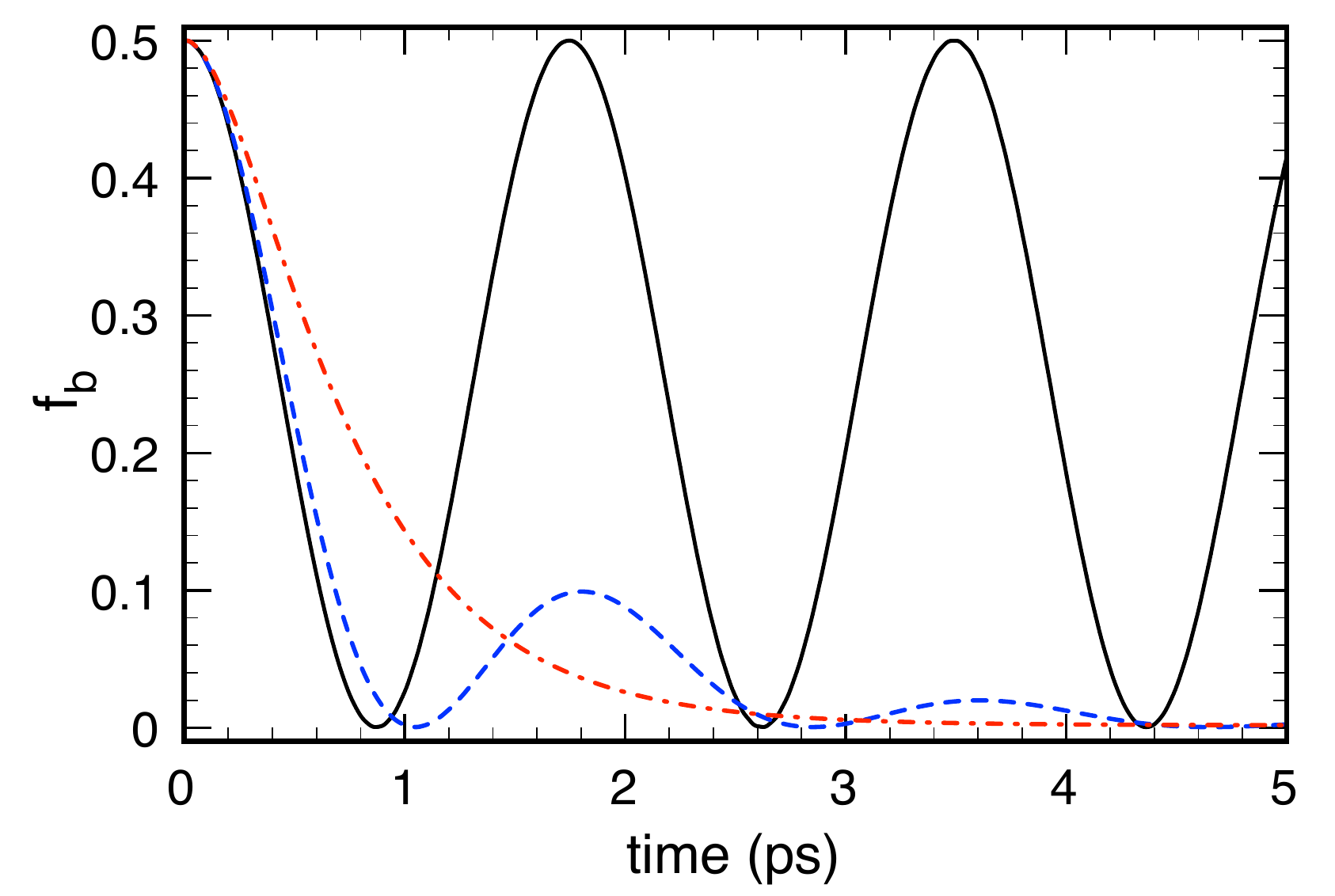}}
\caption{
Time evolution of the excited-level population $f_b$ obtained via a numerical solution of the global density-matrix equation (\ref{EE}) for the same physical parameters considered in Fig.~\ref{Fig3} and for different values of the phonon thermalization rate: 
$\Gamma_{\rm p} = 0$ (solid curve),
$\Gamma_{\rm p} = \overline{\omega}$ (dashed curve), and
$\Gamma_{\rm p} = 5 \overline{\omega}$ (dash-dotted curve).
}
\label{Fig5}
\end{figure}

To better point out the totally unphysical character of the quantum-kinetic results in Figs.~\ref{Fig3} and \ref{Fig4}, we can perform an exact numerical solution of the electron-plus-one phonon system just considered. More specifically, by denoting with $\hat{\rhotot}$ the global (electron-plus-phonon) density-matrix operator, its time evolution can be expressed as
\begin{equation}\label{EE}
\frac{d\hat{\rhotot}}{dt}
=
\frac{1}{\imath\hbar}\,\left[\hat H, \hat{\rhotot}\right]
+
\Gamma_{\rm p}
\left(
\hat b^{ }_{\overline{\mathbf{q}}}
\hat{\rhotot}
\hat b^\dagger_{\overline{\mathbf{q}}}
-
\frac{1}{2} 
\left\{
\hat b^\dagger_{\overline{\mathbf{q}}}
\hat b^{ }_{\overline{\mathbf{q}}}
,
\hat{\rhotot}
\right\}\right)\ ,
\end{equation}
where the first contribution is the usual Liouville-von Neumann term corresponding to the total Hamiltonian $\hat H$, while the second (Lindblad-type) superoperator describes the coupling of our single phonon $\overline{\mathbf{q}}$ with a zero-temperature thermal reservoir in terms of the same phonon relaxation rate $\Gamma_{\rm p}$ introduced in Eq.~(\ref{QKE1}).
Figure \ref{Fig5} shows the time evolution of the excited-level population $f_b$ obtained via a numerical solution of the global density-matrix equation (\ref{EE}) for different values of the phonon thermalization rate $\Gamma_{\rm p}$. In the thermalization-free case ($\Gamma_{\rm p} = 0$) one gets a periodic population/depopulation with frequency $\overline{\omega}$, often referred to as polaronic oscillations; in contrast, for increasing values of the thermalization rate an exponential-like behaviour is progressively established, preserving in any case the initial zero-derivative behaviour typical of exact as well as of quantum-kinetic treatments.
These results are in clear contrast with those in Figs.~\ref{Fig3} and \ref{Fig4}: indeed, in spite of the fact that the latter have been obtained assuming the phonon system in thermal equilibrium ($n^{ }_{\mathbf{q}} \to n^{\circ}_{\mathbf{q}}$), 
in the dispersionless limit the quantum-kinetic result of Fig.~\ref{Fig4} shows a dissipation-free behaviour; 
in contrast, for the very same thermalized-phonon limit, i.e., $\Gamma_{\rm P} \gg \overline{\omega}$, the exact result in Fig.~\ref{Fig5} shows a fully dissipative/incoherent dynamics.

The analysis presented so far shows that, in addition to the totally unphysical negative-value regions of Figs.~\ref{Fig3} and \ref{Fig4}, within the quantum-kinetic model in Eq.~(\ref{QKE2}) a dispersionless phonon mode is totally equivalent to a single phonon mode $\overline{\mathbf{q}}$ and does not induce any dissipation/decoherence on our electronic two-level system; the latter is instead the result of a highly non-trivial interference process involving all different electron and phonon energies. 
Indeed, for the case of a continuous electronic and/or phononic energy band all phonon-assisted density-matrix elements rotate with different frequencies:
$\rhopa^{\mathbf{q}}_{\alpha_1\alpha_2} \propto e^{-\frac{\imath}{\hbar} \left(\epsilon_{\alpha_1}-\epsilon_{\alpha_2}+\epsilon_{\mathbf{q}}\right)}$.
This implies that the time evolution of the density matrix $\rho_{\alpha_1\alpha_2}$ [see first line in Eq.~(\ref{QKE2})] is the result of a purely coherent and reversible interference process among all phonon-assisted density matrices; the situation is similar to the ultrashort temporal decay of the total polarization in a photoexcited semiconductor (see, e.g., \cite{b-Rossi11}) resulting from the coherent superposition of microscopic polarizations rotating with different frequencies (inhomogeneous broadening) and present also in the absence of genuine decoherence processes (homogeneous broadening).
It is however clear that for the case of a 0D electronic system the effectiveness of such interference process requires the presence of a continuum of phonon-assisted density-matrix energies much larger than the typical interstate energy splitting $\epsilon_{\alpha}-\epsilon_{\alpha'}$.
While for the case of the acoustic-phonon mode (see Figs.~\ref{Fig1} and \ref{Fig2}) such a requirement is always fulfilled, the same does not apply to the case of the optical-phonon mode; this is the main origin of the highly unphysical results in Figs.~\ref{Fig3} and~\ref{Fig4}.
 
At this point a few crucial comments are in order.
The physical equivalence between a dispersionless phonon mode and a single phonon $\overline{\mathbf{q}}$ just discussed is based on the assumption of a fully-coherent electron-phonon dynamics; in any realistic nanomaterial, however, the phonon-assisted density-matrix elements $\rhopa^{\mathbf{q}}_{\alpha_1\alpha_2}$ 
are always characterized by a finite lifetime (not included in the present quantum-kinetic treatment) bounded by the phonon thermalization rate $\Gamma_{\rm p}$. 
This explains why for $\Gamma_{\rm p} = 0$ both the quantum-kinetic results of Fig.~\ref{Fig4} and the exact ones in Fig.~\ref{Fig5} show a dissipation-free behavior; however, opposite to the exact result, the quantum-kinetic one is totally unphysical.
In contrast, in the presence of the phonon lifetime just recalled, the exact results in Fig.~\ref{Fig5} exhibit a dissipation/decoherence dynamics; 
this suggests that in realistic nanomaterials an optical-like phonon mode will induce a dissipation/decoherence dynamics closer to the Markov result of Fig.~\ref{Fig3} and thus totally different from its quantum-kinetic counterpart.

\begin{figure}
\centerline{\includegraphics[width=.4\textwidth]{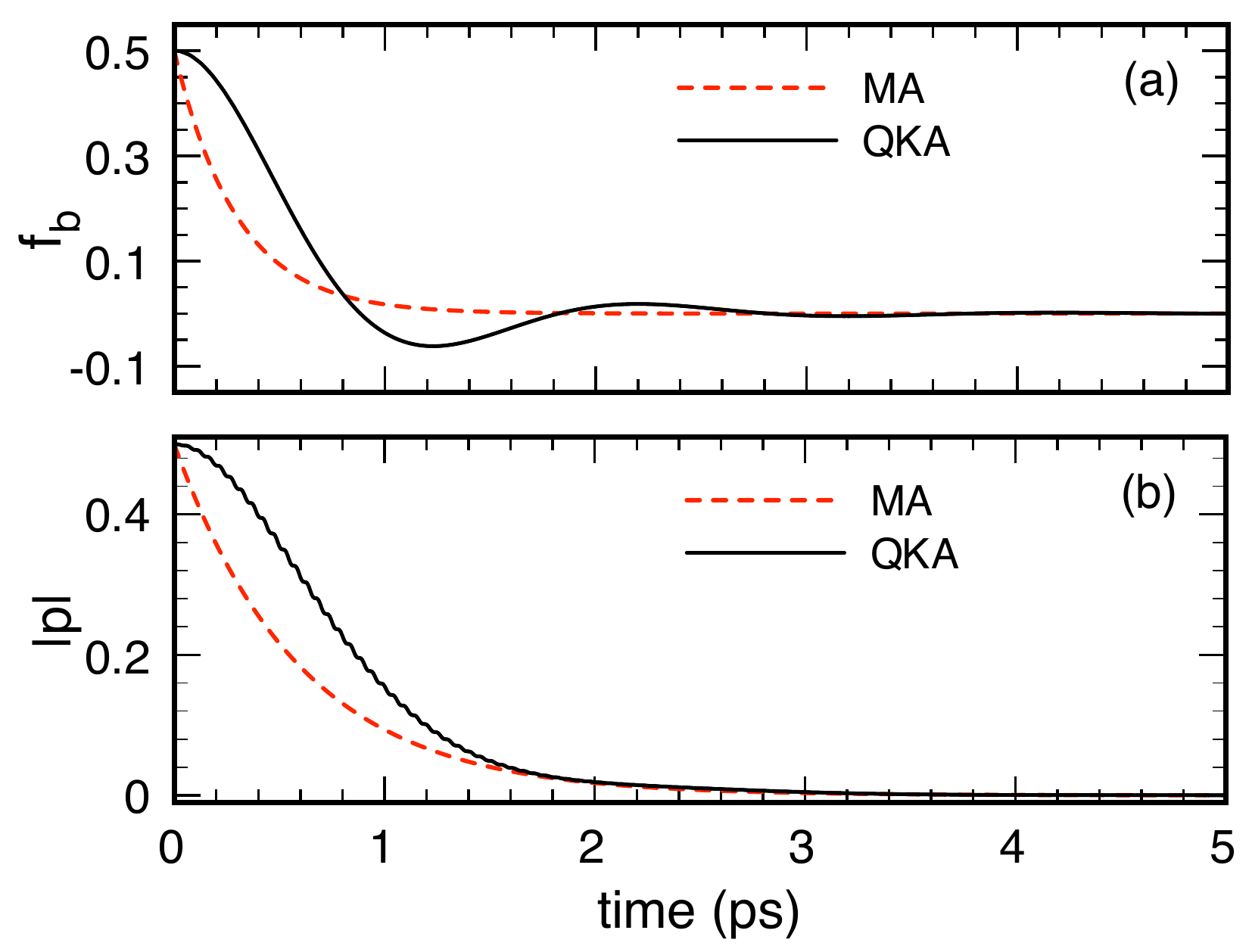}}
\caption{
Same as in Fig.~\ref{Fig3} but in the presence of the carrier-phonon self-energy correction.
}
\label{Fig6}
\end{figure}

It is worth stressing that one of the unphysical features presented so far, namely the occurrence of negative carrier-distribution regions, has already been recognized during the early days of electron-phonon quantum kinetics (see, e.g., \cite{Zimmermann94a} and references therein); a possible strategy to overcome such limitation is the inclusion of a second-order electron-phonon self-energy. 
Within the quantum-kinetic treatment described above, this can be realized extending the correlation expansion up to the third order (see, e.g., \cite{Rossi02b}); 
neglecting energy-band renormalizations (so-called polaronic shifts),
this amounts to add to the (real) single-particle energies $\epsilon_\alpha$, entering the equation of motion for the phonon-assisted density matrix in (\ref{QKE2}), a corresponding (imaginary) lifetime term via the self energy
$\Sigma_\alpha = \epsilon_\alpha + \frac{\imath\hbar}{2} \Gamma_\alpha$.
To check whether such improvement really allows one to overcome the unphysical results reported in Fig.~\ref{Fig3}, we repeat those simulated experiments including the above self-energy correction: for the two-level-system parameters of Fig.~\ref{Fig3} there are no self-energy correction for the ground level $a$ and a finite correction for the excited state $b$ corresponding to a carrier-phonon lifetime $\Gamma^{-1}_b = P^{-1}_{b \to a} = 0.3$\,ps.
Figure \ref{Fig6} shows that, as expected, the inclusion of self-energy terms leads to a significant increase of dissipation and decoherence; nevertheless, we still deal with negative values of $f_b$.
This forces us to conclude that the self-energy improvement just discussed is far from being a general solution to our problem.

An important remark is finally in order.
In the investigation of bulk as well as of quantum-well and quantum-wire structures, whose continuous electronic spectrum allows anyway for a proper treatment of dispersionless phonon modes, the physical limitations discussed so far come out to cause negligible consequences. However, they are expected to have an increasing impact moving from nanomaterials of higher dimensionality down to genuine 0D (i.e., fully discrete-level) systems like, e.g., semiconductor quantum dots, potential candidates for solid-state qubits and quantum logic gates. This occurs in particular in the regime of strong coupling with nearly dispersionless phonons, which is typically the case for GaAs- and GaN-based quantum-dot nanostructures.
For a quantitative investigation of such quantum devices the present two-level modeling needs to be replaced by a more realistic electronic spectrum; nevertheless, all the criticalities discussed so far are expected to affect more refined simulation models as well.

In summary, our analysis has pointed out intrinsic limitations of standard quantum-kinetic approaches in the study of energy dissipation and decoherence in solid-state quantum devices. More specifically, opposite to the positive-definite nature of the Markov treatment in~(\ref{LBE}), the quantum-kinetic scheme in~(\ref{QKE1}) does not ensure/preserve the positive-definite character of the electronic density matrix. However, when the system-environment coupling  becomes  strong and/or the excitation timescale is extremely short, Markovian approaches are known to be unreliable, and memory effects have to be taken into account; for this reason, the derivation of alternative non-Markovian treatments able to ensure the positive-definite character of the single-particle density matrix would be of paramount importance.
To this aim, as already pointed out (see, e.g., \cite{Zimmermann94a} and references therein), a possible way to improve conventional density-matrix treatments is to include higher-order correlation functions, which allows one to incorporate a finite life-time of the phonon-assisted density matrix; indeed, similar treatments have also been adopted within the Green's-function formalism including a corresponding electron-phonon self-energy.
In addition, for few special cases, the very same reliability of the correlation expansion up to higher orders has been tested and compared with exact solutions~\cite{Axt99a,Glassl11a}.
It is however imperative to stress that, while these higher-correlation and/or renormalization schemes may work well for specific systems and simulation parameters, a general quantum-kinetic formalism able to ensure the positive-definite character of our electronic state is still missing.


\end{document}